\documentclass[preprint,showpacs,preprintnumbers,amsmath,amssymb]{revtex4}
\usepackage{graphics,epsf,longtable}
\usepackage{graphicx}  % needed for figures
\usepackage{dcolumn}   % needed for some tables

\begin{document}

\title{Over-the-barrier electron detachment in the hydrogen negative ion}

\author{M. Z. Milo\v sevi\'c}
\affiliation{Faculty of Science and Mathematics, University of
Ni\v s, Vi\v segradska 33, 18000 Ni\v s, Serbia}

\author{N. S. Simonovi\'c}
\affiliation{Institute of Physics, University of Belgrade, P.O.
Box 57, 11001 Belgrade, Serbia}

\begin{abstract}
The electron detachment from the hydrogen negative ion in strong
fields is studied using the two-electron and different
single-electron models within the quasistatic approximation. A
special attention is payed to over-the-barrier regime where the
Stark saddle is suppressed below the lowest energy level. It is
demonstrated that the single-electron description of the lowest
state of ion, that is a good approximation for weak fields, fails
in this and partially in the tunneling regime. The exact lowest
state energies and detachment rates for the ion at different
strengths of the applied field are determined by solving the
eigenvalue problem of the full two-electron Hamiltonian. An
accurate formula for the rate, that is valid in both regimes, is
determined by fitting the exact data to the expression estimated
using single-electron descriptions.
\end{abstract}

\pacs{32.80.Gc, 32.60.+i}

\keywords{hydrogen negative ion; strong field; barrier; tunneling,
electron detachment}

\maketitle

Although it is one of the simplest systems in atomic physics, the
hydrogen negative ion (H$^-$) is still a subject of extensive
experimental and theoretical studies. In early years of quantum
mechanics at the focus of these studies was the ground state of
the free ion. The existence of H$^-$ as a bound system had been
proposed theoretically by Bethe in 1929 \cite{bethe} (see a
historical review of H$^-$ in Ref.~\cite{rau} and an overview in
the context of negative ions in Ref.~\cite{andersen}). Earlier
predictions based on simple perturbational or variational methods
(using, for example, the variational wave function $\psi \sim
\exp[-a(r_1+r_2)]$) had failed, even though these methods were
well suited to predict the most of properties of other members of
the two-electron isoelectronic sequence such as He, Li$^+$,
Be$^{++}$, etc. This is not surprising since the interaction
between electrons in the hydrogen negative ion, unlike to helium
atom and two-electron positive ions, is comparable in magnitude to
that between the nucleus and electrons. As a consequence H$^-$ is
a weakly bound system which has only one bound state -- the ground
state. Its binding energy is $E_\mathrm{B} = 0.7542$\,eV
(0.0277\,a.u.) \cite{AHH,LML,radzig}. A very weak binding and the
absence of a long-range Coulomb attraction for the separated
electron (the atomic residue is the neutral hydrogen atom) results
in the fact that this two-electron system has no singly excited
states.

\begin{figure}
\vspace{.1in}
\includegraphics[width = 3in]{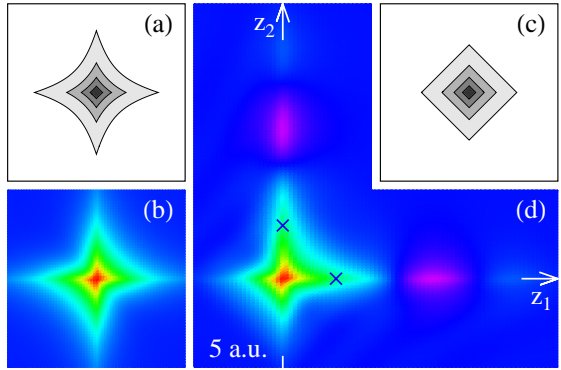}
\caption{(Color online) The $x_1 = y_1 = x_2 = y_2 = 0$ cut of the
real part of: (a) the Chandrasekhar's and (b) the numerically
calculated ground state wave function of H$^-$ (at $F = 0$), (c)
the wave function $\sim \exp[-a(r_1+r_2)]$ representing a state
with equivalent electrons, (d) the numerically calculated lowest
state wave function of H$^-$ in the electric field $F =
0.03$\,a.u. The crosses mark the positions of the barrier saddle
points ($z_\mathrm{sp} \approx 3$\,a.u.) estimated using the
single-electron model (\ref{VCF}). } \label{fig1}
\end{figure}

The wave function and probability amplitude of a weakly bound
system such as H$^-$ can extend beyond the range of the binding
potential itself. As it was recognized by Chandrasekhar more than
70 years ago \cite{chandra} (see also Ref.~\cite{rau}), the ground
state wave function of H$^-$ exhibits a specific radial
correlation between the electrons such that one electron is bound
much closer to the nucleus than the other which is weakly held at
a distance of 4-5 Bohr radii from the nucleus. In contrast to the
wave function with equivalent electrons $\psi \sim
\exp[-a(r_1+r_2)]$ (see Fig.~\ref{fig1}(c)), the Chandrasekhar's
wave function of the form $\psi \sim \exp(-a r_1 -b r_2) + \exp(-a
r_2 -b r_1)$ with the parameters $a = 1.03925$, $b = 0.28309$ (see
Fig.~\ref{fig1}(a)) provides the stability of H$^-$
\cite{chandra}. Regarding the electron detachment processes, such
a configuration suggests a very useful one-electron picture where
the outer electron is weakly (loosely) bound in a short-range
attractive potential well. To a good approximation the potential
acting on the outer electron due to the neutral atom is a sum of a
short-range potential and the polarization term falling off as
$1/r^4$ (see a short overview of the potentials of this type in
the Appendix in Ref.~\cite{gru-sim}). Moreover, since the outer
electron spends much of the time beyond the potential well, it may
be treated even as a free particle subject to boundary conditions
imposed at the nucleus position. This simple model essentially
takes the attraction to be of zero-range and it is in literature
known as the zero-range potential (ZRP, see e.g.
Ref.~\cite{dem-ost}).

Beside strictly theoretical reasons, the earliest interest for
studying H$^-$ came from the atmosphere physics and astrophysics.
The existence of the hydrogen negative ion in the Solar and other
stars photospheres was first discussed in the literature by Wildt
in 1939 \cite{wildt}. In this study it was demonstrated that
photo-absorption properties of H$^-$ might be important for the
opacity of these atmospheres. One of the possible processes which
contribute to the atmospheric absorption coefficient is just the
electron photo-detachment of this ion.

The single-photon detachment cross section for H$^-$ has received
a considerable amount of attention in the past (see
Refs.~\cite{rau,andersen} and references therein). At the
threshold of this process (for one-electron ejection) the residual
hydrogen atom is left in the ground state and no long-range forces
act on the departing electron \cite{BH}. The experimental cross
section is found to be in a good agreement with the Wigner low
that is a feature of short-range potentials \cite{LML}. During the
last two decades, intense lasers have made it possible to observe
effects of multiphoton absorption by atoms and ions, including the
hydrogen negative ion \cite{rau,andersen}. In contrast to the
single-photon case, the multiphoton detachment may occur at the
photon energies $\hbar\omega < E_\mathrm{B}$, but since the
detachment rates in this case are significantly lower, in order to
get a measurable effect one needs much stronger fields.

At larger intensities, however, another mechanism for the electron
detachment arises -- the quantum-mechanical tunneling. A strong
field distorts the potential of atomic residue forming a potential
barrier (Stark saddle) through which the electron can tunnel.
Finally, at a sufficiently strong field the barrier is suppressed
below the energy of the bound state. This regime can be referred
to as over-the-barrier detachment (OBD). The transition from the
multiphoton to the tunnelling regime is governed by the Keldysh
parameter $\gamma = \omega\, (2 m_e E_\mathrm{B})^{1/2}/eF$
\cite{keldysh}, where $F$ is the peak value of the electric
component of electromagnetic field. This parameter characterizes
the degree of adiabaticity of the motion through or over the
barrier: If $\gamma \gg 1$ (high-intensity--long-wavelength limit)
multiphoton processes dominate, whereas for $\gamma \ll 1$
(low-intensity--short-wavelength limit) the tunneling or OBD
mechanism does.

In the second case ($\gamma \ll 1$) the quasistatic description is
a good approximation. It assumes that the electric field changes
slowly enough that a static detachment rate can be calculated for
each instantaneous value of the field. Then the detachment rate
for the alternating field can be obtained by averaging the static
rates over the field period. For this purpose it is sufficient to
use the Hamiltonian (here and thereafter we use the atomic units)
\begin{equation}
H = -\frac{1}{2}(\Delta_1 + \Delta_2) - \frac{1}{r_1} -
\frac{1}{r_2} + \frac{1}{r_{12}} - F(z_1 + z_2), \label{ham2e}
\end{equation}
describing the dynamics of two electrons of H$^-$ in a static
electric field $F$. Here $\mathbf{r}_i$ is the $i$-th electron's
position and $r_{12} = |\mathbf{r}_1 - \mathbf{r}_2|$. Due to
presence of the barrier all eigenstates of (\ref{ham2e}) have the
resonant character when $F \neq 0$, including the lowest which is
an exact bound state for $F = 0$. The width $\Gamma$ of the lowest
state determines the electron detachment rate $w(F) =
\Gamma(F)/\hbar$ (hereafter we set $\hbar = 1$).

The eigenstates of (\ref{ham2e}) are calculated numerically using
the complex rotation method \cite{reinhardt,buchleitner}. The
calculations are performed in the basis whose elements are the
symmetrized products of Sturmian functions \cite{avery} for each
electron. Fig.~\ref{fig1}(b,d) shows 2D cuts of the ground state
of H$^-$ ($F = 0$) and the lowest state of H$^-$ in the field of
strength $F = 0.03$\,a.u., respectively. By comparing the parts
(a) and (b) of the same figure, one can see that the
Chandrasekhar's wave function is indeed a good approximation for
the ground state of H$^-$. A small difference is due to the lack
of angular correlations in the approximate wave function. The
outgoing waves of the wave function shown in Fig.~\ref{fig1}(d),
representing the (single-electron) escape channels for the first
and for the second electron, clearly demonstrate the resonant
character of this state.

\begin{figure}
\includegraphics[width = 3in]{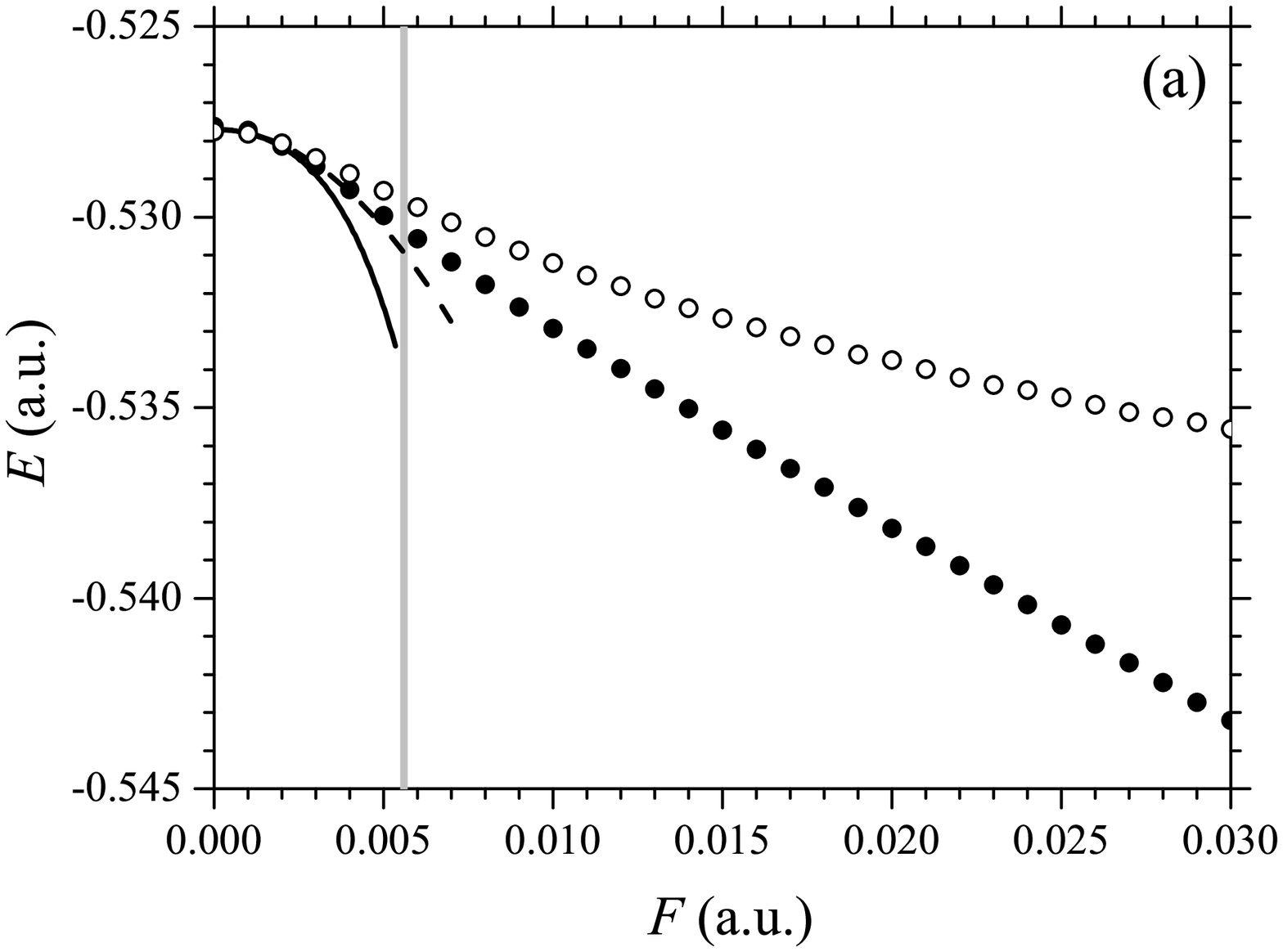}
\includegraphics[width = 3in]{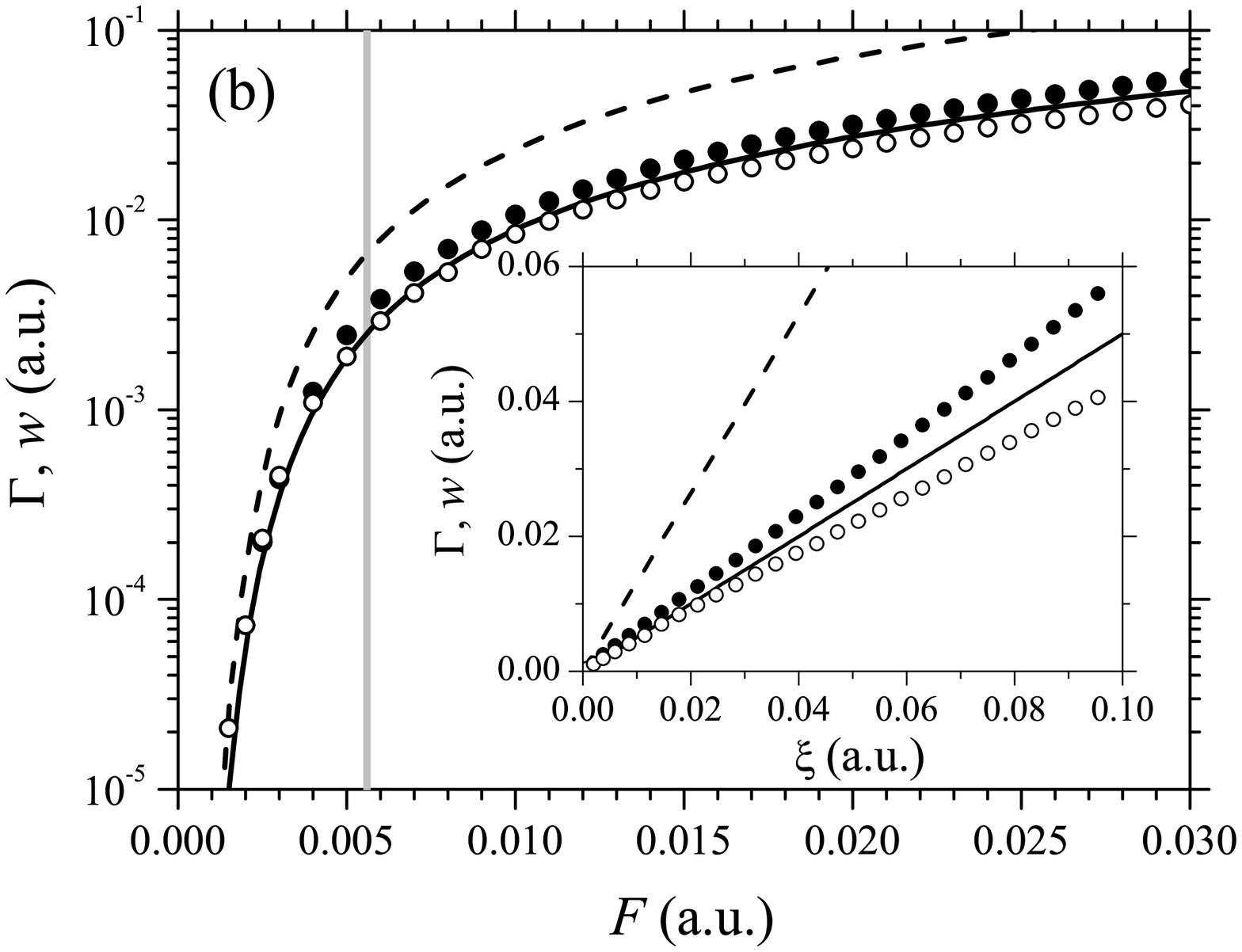}
\caption{(a) The lowest state energy $E$ and (b) width $\Gamma$
(detachment rate $w$) of the hydrogen negative ion as functions of
the strength of applied electric field $F$. The full and open
circles denote the results obtained numerically using the two- and
the single-electron CF (Eq.~(\ref{VCF})) model, respectively. The
dashed and full lines in part (a) show the lowest state energy
$E(F)$ estimated using the second and the fourth order Stark shift
formula, respectively. The dashed and full lines in part (b) show
the rate $w(F)$ given by the PPT and the ZRP theory
(Eq.~(\ref{det-rate}) with $C_\kappa^2 = 1.32$ and $0.5$),
respectively. The inset shows the same rates as functions of the
variable $\xi$ (see text). Vertical gray lines separate the
tunneling from OBD regime.} \label{fig2}
\end{figure}

The lowest state energies and widths (electron detachment rates)
of H$^-$ at different values of the applied electric field
determined numerically using the two-electron model (\ref{ham2e})
are presented in Table~\ref{table} and in Fig.~\ref{fig2} together
with the results obtained using other approaches.
Fig.~\ref{fig2}(a) shows that for weak fields the lowest state
energy $E(F)$ (numerical data) decreases by increasing the field
strength according to the Stark shift expansion formula $\Delta E
\equiv E(F) - E(0) = -\alpha F^2/2! - \gamma F^4/4! - \cdots$.
Here $E(0) = -0.5 + E_\mathrm{B} = -0.5277$\,a.u. is the ground
state energy of the free ion, whereas $\alpha = 206$ and $\gamma =
8.03 \cdot 10^7$ are the corresponding values of the dipole
polarizability and the second dipole hyperpolarizability
\cite{radzig,pipin}. At stronger fields ($F > 0.01$), however, the
lowest state energy depends on the field strength almost linearly.
The same figure shows the results obtained using a single-electron
model that will be discussed later.

As mentioned in the introductory part, the configuration of the
ground state of H$^-$ suggests a one-electron description where
the outer (loosely bound) electron moves in a short-range
potential $V(\mathbf{r})$ describing the attraction by the neutral
atomic residue. Then, in the presence of a (quasi)static electric
field $F$ the outer electron may be considered as moving in the
total potential $V_\mathrm{tot} = V(\mathbf{r}) - F z$.
$V(\mathbf{r})$ is usually calibrated to give the value $-E_B$ for
the lowest energy level $\epsilon(F)$ at $F = 0$. When $F \neq 0$
the total potential has a potential barrier that explains the
resonant character of states. The saddle point of the barrier is
located at the z-axis. Its position $\mathbf{r}_\mathrm{sp} =
(0,0,z_\mathrm{sp})$ and hight $V_\mathrm{sp} =
V_\mathrm{tot}(\mathbf{r}_\mathrm{sp};F)$ depend on the field
strength $F$ and can be determined from the rule $(\partial
V_\mathrm{tot}/\partial z)_{x = y = 0} = 0$. The field strength
$F_\mathrm{S}$ that separates the tunneling and OBD regimes is
defined by the condition $\epsilon(F_\mathrm{S}) =
V_\mathrm{sp}(F_\mathrm{S})$. Note that these values may vary by
changing the model for $V(\mathbf{r})$.

%\begingroup
\squeezetable
\begin{table}
\caption{The lowest state energies $E$ and widths $\Gamma$ of the
hydrogen negative ion at different strengths of applied electric
field $F$, obtained by the complex rotation method within the
single- and two-electron pictures. The single-electron
calculations are performed using the CF potential (\ref{VCF}).}
\label{table}
%\begin{ruledtabular}
%\begin{tabular}{ccccc}
%\\[-2.5ex]%\\[-2.5ex]
\begin{tabular}{@{\extracolsep{1ex}}ccccc}
\hline\hline
\\[-2ex]
&\multicolumn{2}{c}{two-electron
model}&\multicolumn{2}{c}{single-electron model}
\\[.2ex]
\hline
\\[-2ex]
$\quad F\quad$&$-E$&$\Gamma$&$-E$&$\Gamma$
\\
\hline
\\[-2ex]
0       & 0.52763  & 0 & 0.52775  & 0         \\
0.001   & 0.52773  & - & 0.52782  & -         \\
0.002   & 0.52814  & - & 0.52806  &$7.330\!\cdot\!\! 10^{-5}$  \\
0.003   & 0.52867  & $4.310\!\cdot\!\! 10^{-4}$ & 0.52846  &$4.511\!\cdot\!\! 10^{-4}$  \\
0.004   & 0.52928  & $1.247\!\cdot\!\! 10^{-3}$ & 0.52887  &$1.091\!\cdot\!\! 10^{-3}$  \\
0.005   & 0.52997  & $2.475\!\cdot\!\! 10^{-3}$ & 0.52931  &$1.913\!\cdot\!\! 10^{-3}$  \\
0.006   & 0.53057  & $3.845\!\cdot\!\! 10^{-3}$ & 0.52974  &$2.933\!\cdot\!\! 10^{-3}$  \\
0.007   & 0.53118  & $5.369\!\cdot\!\! 10^{-3}$ & 0.53014  &$4.122\!\cdot\!\! 10^{-3}$  \\
0.008   & 0.53177  & $7.022\!\cdot\!\! 10^{-3}$ & 0.53053  &$5.335\!\cdot\!\! 10^{-3}$  \\
0.009   & 0.53236  & $8.789\!\cdot\!\! 10^{-3}$ & 0.53088  &$7.034\!\cdot\!\! 10^{-3}$  \\
0.010   & 0.53293  & 0.01066 & 0.53121  &$8.440\!\cdot\!\! 10^{-3}$  \\
0.011   & 0.53347  & 0.01258 & 0.53153  &$9.872\!\cdot\!\! 10^{-3}$  \\
0.012   & 0.53397  & 0.01451 & 0.53182  & 0.01132  \\
0.013   & 0.53451  & 0.01654 & 0.53214  & 0.01284  \\
0.014   & 0.53503  & 0.01861 & 0.53240  & 0.01443  \\
0.015   & 0.53559  & 0.02078 & 0.53265  & 0.01595  \\
0.016   & 0.53609  & 0.02291 & 0.53289  & 0.01750  \\
0.017   & 0.53660  & 0.02505 & 0.53313  & 0.01894  \\
0.018   & 0.53708  & 0.02730 & 0.53336  & 0.02065  \\
0.019   & 0.53762  & 0.02956 & 0.53360  & 0.02227  \\
0.020   & 0.53817  & 0.03186 & 0.53376  & 0.02394  \\
0.021   & 0.53864  & 0.03414 & 0.53399  & 0.02562  \\
0.022   & 0.53915  & 0.03647 & 0.53422  & 0.02720  \\
0.023   & 0.53965  & 0.03883 & 0.53441  & 0.02883  \\
0.024   & 0.54016  & 0.04122 & 0.53454  & 0.03066  \\
0.025   & 0.54071  & 0.04362 & 0.53473  & 0.03229  \\
0.026   & 0.54120  & 0.04606 & 0.53492  & 0.03395  \\
0.027   & 0.54170  & 0.04848 & 0.53512  & 0.03563  \\
0.028   & 0.54222  & 0.05097 & 0.53525  & 0.03733  \\
0.029   & 0.54274  & 0.05349 & 0.53538  & 0.03900  \\
0.030   & 0.54320  & 0.05599 & 0.53556  & 0.04062  \\
%[-.5ex]
\hline\hline
\end{tabular}
%\end{ruledtabular}
\end{table}
%\endgroup

The first among the single-electron approaches we will consult is
the Perelomov-Popov-Terent'ev (PPT) theory \cite{PPT}. It is based
on the quasistatic approximation and the assumption that most
atoms are nearly hydrogenic, the difference being a small quantum
defect that changes the quantum numbers to noninteger effective
values. In the case of negative ions, however, the atomic residue
is neutral ($Z = 0$) and the effective principal quantum number
$n^* = Z/\kappa$, where $\kappa = (2E_\mathrm{B})^{1/2}$, is equal
to zero. Then the static-field tunneling rate formula for negative
ions in the ground state reduces to
\begin{equation}
w = C_\kappa^2 \frac{F}{\kappa}
\exp\!\bigg(\!-\frac{2\kappa^3}{3F}\bigg). \label{det-rate}
\end{equation}
The coefficient in the pre-exponential factor for H$^-$ determined
from Hartree-Fock calculations has the value $C_\kappa = 1.15$
\cite{popov}. Fig.~\ref{fig2}(b) shows that, although H$^-$ does
not belong to the class of hydrogenic atoms, the PPT rate formula
exhibits a qualitative agreement with the numerical results.

It should be mentioned that the Ammosov-Delone-Krainov (ADK)
theory \cite{ADK} gives the same (genaral) formula for tunneling
rates as the PPT theory, but in addition it provides an explicit
expression for $C_\kappa$. This expression, however, is not useful
in the case when $n^* = 0$. The ADK theory, on the other hand,
accurately predicts tunneling rates in experiments with atomic
ionization in strong fields \cite{augst,xiong} and shows good
agreement with available exact numerical results (for H and He see
\cite{scrinzi,themelis}). Even for atoms with low ionization
potentials like alkali metals, using a correction which accounts
for the Stark shift, the ADK tunneling rates agree well with
numerical results \cite{MS}. However, the non-Coulomb character of
interaction between the neutral atomic residue and outer electron
raise the question of applicability of the PPT or a similar theory
to negative ions.

Regarding the latest discussion we can expect a better agreement
between the single- and the two-electron approach if in the former
we apply a short-range potential. As mentioned above, the simplest
short-range potential that can be used to describe the dynamics of
a weakly bound electron in negative ions is the ZRP:
$V(\mathbf{r}) = -a\delta(\mathbf{r})$ ($a > 0$). This potential
supports only one bound state whose wave function has the form
$\psi(\mathbf{r}) \sim \exp(-\kappa r)/r$, where $\kappa \equiv
(2E_\mathrm{B})^{1/2} = a$ \cite{dem-ost}. The eigenvalue problem
of the single-electron Hamiltonian with $V_\mathrm{tot} =
-a\delta(\mathbf{r}) - Fz$ admits for weak fields a solution in a
closed analytical form \cite{dem-ost}. The lowest state energies
and widths are represented by the Stark shift expansion with the
polarizability $\alpha = 1/(16E_\mathrm{B}^2) \approx 81.5$ and by
Eq.~(\ref{det-rate}) with $C_\kappa^2 = 1/2$. Hence, the PPT and
ZRP rate formulae differ only by the value of constant $C_\kappa$.

The exact value of constant $C_\kappa$ can be obtained by fitting
the numerical results obtained applying the two-electron model.
For this purpose we express the rate in terms of the variable $\xi
= (F/\kappa) \exp(-2\kappa^3/(3F))$. Then the rate formula
(\ref{det-rate}) reduces to the linear dependence $w = C_\kappa^2
\xi$. It is found that the numerical data fits well to
Eq.~(\ref{det-rate}) for $C_\kappa^2 = 0.585 \pm 0.010$ (see the
inset in Fig.~\ref{fig2}(b)).

Finally we consider the single-electron model for H$^-$ where the
loosely bound electron moves in an effective potential that is the
sum of a short-range potential and the polarization term. A widely
used potential of this type is the Cohen-Fiorentini (CF) potential
\cite{CF}
\begin{equation}
V = -\bigg(1+\frac{1}{r}\bigg) e^{-2r} -
\frac{\alpha_\mathrm{H}}{2r^4}\, e^{-r_0^2/r^2}, \label{VCF}
\end{equation}
where $\alpha_\mathrm{H} = 9/2$ is the polarizability of the
hydrogen atom. The parameter $r_0 = 1.6$ is chosen by the
condition that the potential (\ref{VCF}) has a single bound state
with the correct binding energy. The lowest state energies [$E(F)
= \epsilon(F) - 0.5$\,a.u.] and widths of the H$^-$ in
(quasi)static electric field, obtained using the CF potential, are
shown in Table~\ref{table} and Fig.~\ref{fig2}. The calculations
were performed using the complex rotation method
\cite{reinhardt,buchleitner} and Sturmian basis \cite{avery}.

At low values of $F$ the energies obtained by the latest model
approximately agree with the two-electron (exact) results (see
Fig.~\ref{fig2}(a)). At stronger fields, however, the difference
between these results increases, particularly in the OBD area. The
value of $F$ that separates the tunneling and OBD regimes obtained
using the potential (\ref{VCF}) is $F_\mathrm{S} = 0.0056 \pm
0.0001$. For $F > 2F_\mathrm{S}$ the exact Stark shift $\Delta
E_\mathrm{2e}$ is approximately two times larger than $\Delta
E_\mathrm{1e}$ obtained using the single-electron approach (the
uncertainty in $F_\mathrm{S}$ is due to this difference). The
rates determined using the CF single-electron model agree with the
two-electron results approximately for $F < F_\mathrm{S}/2$, see
Fig.~\ref{fig2}(b). Otherwise the single-electron calculations
underestimate the two-electron results (for about 30$\%$ in the
OBD regime).

These differences indicate that the single-electron picture is not
valid at stronger fields. At the field strengths $F \sim
F_\mathrm{S}$ the potential barrier is suppressed enough that the
lowest state cannot be treated as bound even approximately. In
this case the Chandrasekhar's concept of outer electron is not
adequate because a significant part of the probability
distribution lies at the outer side of barrier ($z >
z_\mathrm{sp}$). In other words the 'outer' electron becomes the
'outgoing' electron. Simultaneously, the form of the two-electron
wave function in the inner region ($z < z_\mathrm{sp}$) becomes
more similar to that for equivalent electrons (see
Fig.~\ref{fig1}(c,d)), that explains the failure of
single-electron approach (particularly for energies). The ratio
$\Delta E_\mathrm{2e}/\Delta E_\mathrm{1e} \approx 2$ for $F \gg
E_\mathrm{S}$ may be explained by the fact that in the states of
this form the shift $\Delta E_\mathrm{2e}$ includes the
contributions of both electrons.

In conclusion, the single-electron description of the lowest state
of H$^-$, that is a good approximation in the field-free and
low-field cases, fails in OBD and partially in the tunneling
regime. This is important to know because single-electron models
are often used to study negative ions in strong fields. We
determined the exact lowest state energies and detachment rates
for H$^-$ at different strengths of the applied (quasi)static
field by solving the eigenvalue problem of the full two-electron
Hamiltonian. The PPT and ZRP theories lead to the same rate
formula, but with different values of the constant in the
pre-exponential factor. The accurate value of the constant is
obtained by fitting the numerical results determined using the
two-electron model.

This work is supported by the COST Action CM1204 (XLIC). N.~S.~S.
acknowledges support by the Ministry of education, science and
technological development of Republic or Serbia under Project
171020.

\end{document}